\documentclass[aps]{revtex4-2}

\usepackage{amsmath,amssymb}
\usepackage{graphics,graphicx}
\usepackage{dcolumn,bm}
\usepackage{psfrag}
\usepackage{color}
\usepackage{mathrsfs}
\usepackage{graphics,graphicx}
\usepackage{enumitem}
\DeclareMathOperator{\erf}{erf}

\usepackage{blkarray}

\begin{document}

\title{On interactive anisotropic walks in two dimensions generated from a  three state opinion dynamics model}

\author{Surajit Saha$^{1}$, Parongama Sen$^{2}$}%

\address{$^{1}$ Department of Physics, University of Calcutta,
92 Acharya Prafulla Chandra Road, Kolkata 700009, India.\\
$^{2}$ Department of Physics, University of Calcutta,
92 Acharya Prafulla Chandra Road, Kolkata 700009, India.}

\keywords{Markovian and non-Markovian processes, phase transitions, random walk, probability distribution, crossover, urn model, ising model,opinion dynamics, sociophysics, critical exponent, bivariate distribution }

\email{psphy@caluniv.ac.in}

\begin{abstract}

A system of interacting walkers is considered in a two-dimensional hypothetical space, where the dynamics of each walker are governed by the opinion states of  the agents of a fully connected three-state opinion dynamics model. Such walks, studied in  different models of statistical physics, are usually considered in one-dimensional virtual spaces. Here, the mapping is done in such a way that the walk is directed along the $y$ axis while it can move either way along the $x$ axis. 
The walk  shows that there are three distinct regions as the noise parameter, 
responsible for driving a continuous phase transition in the model,
is 
varied. 
In absence of any noise, the scaling properties and the form of the distribution along either axis do not follow
any conventional form. 
For any finite noise below  the critical point the bivariate distribution of the displacements 
is found to be a modified biased Gaussian function while above it,
only the marginal distribution along one direction is Gaussian. 
The marginal probability distributions can be extracted and the scaling forms of different quantities, showing power law behaviour, are obtained. The directed nature of the walk is reflected in the marginal distributions as well as in the exponents.

\end{abstract}

\pacs{}

\maketitle

\section{Introduction}

Exploring the dynamics of a physical system often involves a strategy of mapping it into an alternative realm, where new types of interacting objects or pseudo-objects are involved. Several such examples are already documented in the literature. The well-known connection between the zero-temperature spin coarsening dynamics in a one-dimensional Ising–Glauber model \cite{priv,derr1,derr2,derr3} and the diffusive motion of domain walls, which undergo annihilation upon collision, is a  notable example that has attracted significant attention. For the voter model \cite{ligg}, one can conceive of a system of coalescing walkers which is equivalent to
the dynamics of the agents \cite{ligg,krap,howard} in any dimension. Some of the other notable examples that have received  attention recently include the connection between Polya-type urn models and discrete-time random walks with memory \cite{surajit2,pol}, as well as the connection between certain coupled oscillators and quantum algorithms \cite{gog}. \\

Any dynamic (stochastic) process taking place at discrete time steps can be regarded as a walk in a virtual space, where the displacements of the walkers correspond to the dynamical variable at that time. 
To be precise, the position vector of the $i$th walker in the virtual space can be written as
\begin{equation}
{\mathbf r}_i (t+1) = {\mathbf r}_i (t) + {\mathbf \Xi}_i (t+1), 
\end{equation} where ${\mathbf \Xi}_i$ are given in terms of the relevant variables (e.g., spin, opinion etc.) related to the original model. For systems with many degrees of freedom interacting with each other, the resultant walks become correlated indirectly. Such walks,
 considered earlier in quite a few studies \cite{godre3,godre4,acps,pratik,sanchari,surajit1,kb2024}, have been shown to  carry the signature of the phase transitions, if any, and can be related to the persistence properties of the system. It is easy to define such a walk with displacements $\pm 1$ corresponding to Ising spins or binary opinion models in a virtual one-dimensional space. When models with more than two states are considered, such walks can still be defined. In an earlier work by the present authors \cite{surajit1}, the one-dimensional walks corresponding to the Biswas-Sen-Chatterjee  (BChS henceforth) model \cite{soumya2012} of opinion formation with three opinion states $\pm 1, 0$ were generated.  It was assumed that for the zero state, the walker does not make any movement, while  the $\pm 1$ states corresponded to dispacements along a one dimensional line in opposite direction. However, this suppresses the role and effect  of the ``zero'' states of the system and the position of the walkers will be independent of the number of times such states have been attained.

 In this work, we have considered a two-dimensional (2D) virtual walk  corresponding to the  three-state BChS model on a fully connected network which takes care of the zero opinion state as well. Note that the dimension of the space in which the virtual walk takes place is unrelated to the spatial dimension of the original system. The walks generated are by definition  istropic  along the X axis and directed along the Y axis. 
Directed anisotropic walks in two dimensions have been considered before as an independent problem 
\cite{direct} and were shown to manifest ballistic behavior at long time scales. 

The primary interest is in the distribution of the displacements in the virtual space.
The results in \cite{surajit1} indicated the presence of  two biased Gaussian distributions, 
centered symmetrically at positive and negative values,   below the critical point and a Gaussian centered around zero above it. The studies  conducted  were mainly done close to the critical point. However, the study of the dynamics of the BChS model in two dimensions yields a number of non-intuitive results even in absence of noise \cite{sudip}. In the present study therefore, 
we allow the noise parameter  to vary over a larger range. The deviation from a biased Gaussian becomes evident in absence of any noise,  the detailed results are presented later in the paper. Usually, the distribution below the critical point is 
non-Gaussian, so any departure from the biased Gaussian form 
has been explored more carefully in this region in the present work. 

All information of the one dimensional walk considered in \cite{surajit1}  are retained
even when the dynamics are regarded as a two dimensional virtual walk.  This is because  one can recover the 
one dimensional distribution along  the X axis as a marginal distribution  of the 2D probability density expressed as a function of $x$ and $y$ coordinates. In addition, one can also obtain a radial distribution, which, although one
dimensional in principle, will not be identical to the former. 
Virtual walks are specially useful to study persistence probabilities. The two dimensional walk is defined in such a way that the persistence probability with respect to any opinion can be extracted easily.

 It has been observed earlier that new exponents can be associated purely with the walk features.  One of the objectives in this study is to explore whether any other distinct exponents are associated with the different marginal distributions defined from the 2D walk.  This may add some more insights into the dynamics at the microscopic level. It is also expected to be useful as one can compare  the  results of the three state opinion dynamics model with those in  other  three or multistate models, for example the $S=1$ spin models \cite{soc_rmp,galam_book,ps}, in subsequent studies. 
It has been further pointed out that the opinion dynamics model can be mapped to an urn model so that the present analysis has multiple applications.

\section{The  model  and definition of the virtual walks in two dimensions }

In two dimensions, in general, we have a walk defined by 
\begin{eqnarray}{}
  x(t+1)  = & x(t) + \xi (t+1)  \nonumber \\
  y(t+1) =   & y(t) + \eta (t+1),
  \label{walk1}
\end{eqnarray}
where $\xi$ and $\eta$ are related to the  state of the agent in the  opinion dynamics model. 

 In the BChS model, the opinion of the $i$th individual, $o_i(t)$,  is updated at time $t$ following an interaction with a randomly selected neighbouring agent $j$  in the following manner:

\begin{equation}
o_i(t+1)=o_i(t)+\mu_{ij}o_{j}(t). 
\label{op}
\end{equation}
Here we take $o_i = 0,\pm1$ that may represent the support for parties with different ideologies, e.g., left, central and right, or in a two-party contest, the zero opinions can indicate neutral agents or abstainers. The opinion value is bounded, if it becomes higher (lower) than $+1$ $(-1)$ then it is made equal to $+1$ $(-1)$. The average opinion  $O=\frac{\mid\sum_i o_i \mid}{N}$ can be regarded  
as the order parameter; note that the zero opinions do not contribute in this measure. 

This model has been studied in various contexts and on different topologies \cite{review1}; we consider it on a fully connected graph.  $\mu_{ij}$ is the interaction parameter representing the influence of the  $j$th agent  on the $i$th individual. It can take values $\pm 1$;  a negative value is taken with probability  $p$, the only parameter in the model. 
The above mean-field model can be  solved yielding an 
order-disorder phase transition at $p = p_c = 0.25$, with Ising-like criticality \cite{soumya2012}. 

\begin{figure}
\includegraphics[width=8cm]{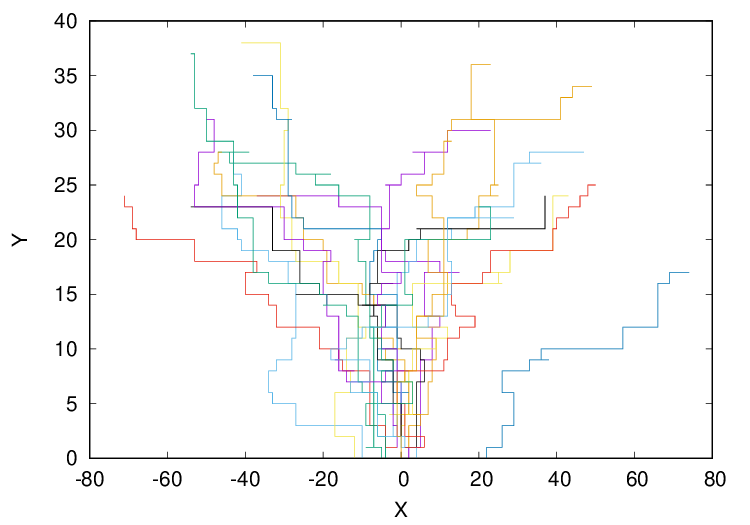}
\caption{A snapshot of the trajectories of 20 walkers in the  XY plane for $p=0.19$ and $t=100$. The walkers, directed along the Y axis, and traversing in both directions along the X axis, carry microscopic information about the system.}
\label{snap}
\end{figure}

In the present two-dimensional walk, the variables $\xi(t)$ and $\eta(t)$ (Eq. \ref{walk1}) of the $i$-th walker, associated with the $i$th agent,  at time step $t$ are determined by the following equations
\begin{equation}
\xi (t)= o_i(t)
\label{dynamics1}
\end{equation}
\begin{equation}
\eta(t)=(1-\lvert o_i(t) \rvert).
\label{dynamics2}
\end{equation}
Thus at each step the i-th walker can perform one of the three following actions: it can move to the nearest-neighbour site to its right or left (movement along $X$ axis) or upwards (movement along $Y$ axis)  corresponding to the opinion value of the i-th individual, i.e. $1$, $-1$ or $0$ respectively. 
Snapshots of the trajectories of the walks of some agents is shown in Fig. \ref{snap} up to a certain time with $p < p_c$.

The equivalent urn model of the above mean field model has been 
discussed in the Appendix \ref{appendix:urn}.

\section{Results}
We performed numerical simulations of the kinetic exchange model for opinion dynamics on a fully connected graph with $N$ nodes. The initial configuration was entirely random, i.e. at time $t=0$, the number of individuals holding opinions of $0$, $+1$, and $-1$ was evenly distributed, with each opinion being assigned to $N/3$ individuals. One Monte Carlo Step (MCS) consists of $N$ updates. During each update, two distinct individuals are selected at random, and the opinion of the first individual is modified based on the rules specified in Eq. \ref{op}. The
maximum system size simulated was $N = 2000$ and the maximum number of configurations over which averaging had
been done was 200000.

\subsection{Bivariate and marginal distributions}

\begin{figure}
  \centering

\hspace{0.4cm}
  \begin{minipage}{0.5\textwidth}
  \includegraphics[width=.9\linewidth,height=140pt]{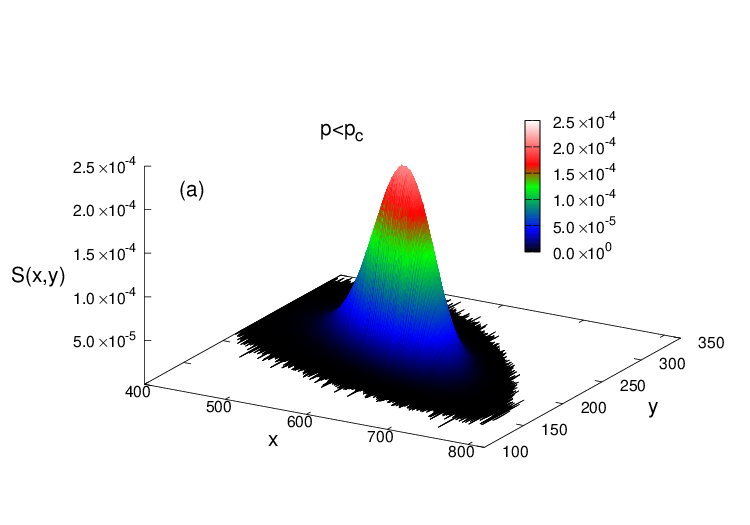} 
\end{minipage}
\vspace{-0.2cm}
\hspace{-9pt}
\begin{minipage}{0.5\textwidth}
   \includegraphics[width=.9\linewidth,height=140pt]{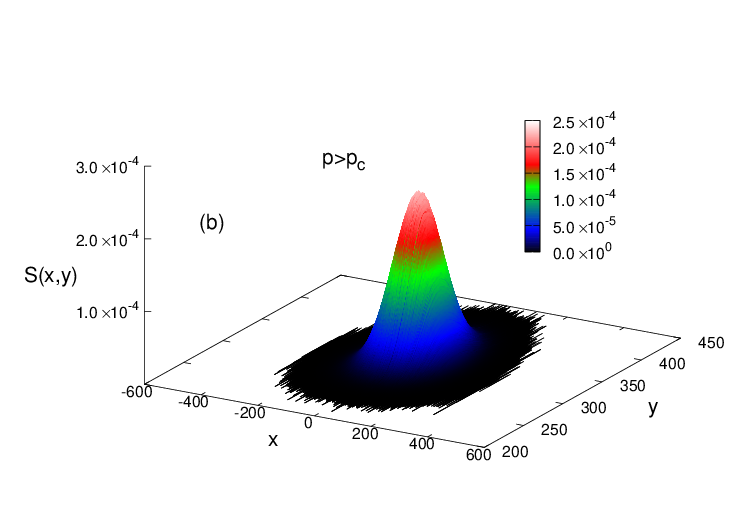} 
\end{minipage}
  \caption{(a) $S(x,y)$ for $p < p_c$ obtained from simulation data along positive X axis for $p=0.19$. The detailed form of the fitted function Eq-\ref{mg} has been discussed in Appendix \ref{appendix:qq}.   (b) $S(x,y)$ for $p = 0.3 >p_c$. Both datasets are   obtained from $N=1000$ at time $ t=1000$.}
  \label{bv}
\end{figure}

At $t=0$, the position of all the walkers are assumed to be $\{0,0\}$. 
We denote by $S(x,y,t)$, the probability that a walker has reached the position $x,y$  at time $t$; in general $S(x,y,t) = S(-x,y,t)$.
The bivariate probability density function of the 2D virtual walk, $S(x,y,t)$, 
is estimated from the numerical simulation. As an example, the data are shown in Fig. \ref{bv}  for  both $p < p_c$ and $p > p_c$;  data for $x> 0$ only are  shown  for the former. 

Marginal probability densities of the bivariate distribution functions along the X and Y axis  can be defined as,
\begin{equation}
P_x(x,t)=\sum_{y=0}^{\infty}S(x,y,t)
\label{margx}
\end{equation}
and

\begin{equation}
P_y(y,t)=\sum_{x=0}^{\infty}S(x,y,t)=\frac{1}{2}\sum_{x=-\infty}^{\infty}S(x,y,t)
\label{margy}
\end{equation}

In the last equation, we have utilised the symmetry of the bivariate distribution $S(x,y,t)$ along the X-axis and defined $P_y(y,t)$ for positive values of $x$ only.   The symmetry along the X-axis occurs due to the fact that a global change of opinions, $o_i\rightarrow -o_i$, keeps Eq. \ref{op} unchanged. 

Angular marginalization refers to marginalizing the bivariate distribution to yield 1-D distributions over radius  $r$ which is the distance from the origin. Radial  marginalized distribution  of $S(x,y,t)$ can be defined as follows
\begin{equation}
P_r(r,t)=\sum_{x=0}^{\infty}\sum_{y=0}^{\infty}\delta_{r,\lfloor\sqrt{x^2+y^2}\rfloor} S(x,y,t),
\label{margr}
\end{equation}
where also the symmetry $x \to -x$ has been used.
It maybe noted here that unlike $x$ and $y$, $r$ can assume non-integer and irrational values. 

We have found that for $0<p<p_c$, the bivariate probability density function can be fitted to the following modified bivariate normal distribution, here for brevity, we suppress the argument $t$ (see Appendix \ref{appendix:qq})
\begin{equation}
S(x,y)=H(x,y)e^{-(m_1(x-\mu_1)^2+m_2(y-\mu_2)^2+m_3(x-\mu_1)(y-\mu_2))},
\label{mg}
\end{equation}
where for a particular $p$, the coefficients $m_1$, $m_2$, $m_3$ are constants and $\mu_1$, $\mu_2$ depend on time. The form of $H(x,y)$ has been discussed in Appendix \ref{appendix:qq}.  Eq. \ref{mg} shows that $S(x,y)$ deviates from a simple biased Gaussian form. $S(x,y)$ for $p>p_c$    shown in Fig. \ref{bv}b  cannot be fitted well with a bivariate normal distribution either (see  Appendix \ref{appendix:qq}).\\

The shape of the distribution (Fig. \ref{bv}a) for $p<p_c$ clearly shows that the correlation of $x$ and $y$, the displacements along the two axes, is negative.  One can define correlation coefficient $\rho_{xy}=\frac{Cov(x,y)}{\sigma_x\sigma_y}$, where $cov(x,y)=\langle xy\rangle-\langle x\rangle\langle y\rangle$ and  $\sigma_x$,$\sigma_y$ are standard deviations. An increment $+1$ or $-1$ of walker position along the X axis corresponds to zero increment along the Y axis and an increment of $+1$ along Y axis corresponds to zero increment along X.  This is the reason why the correlation should be negative.  We have studied the behavior of $\rho_{xy}$ as a function of $p$. While for $p>p_c$ it is nearly zero, below $p_c$, it follows a behavior 
\begin{equation}
    \rho_{xy} \propto -(p_c-p)^{0.37}.
\end{equation}
 Hence we find, from the 2D walk, that apart from the order parameter, there is
another quantity that vanishes above the critical point and shows power law behavior as a function of $p_c-p$ below it. However, the known value of the critical exponent 
associated with the order parameter is 0.5, clearly different from that of $\rho$.
\begin{figure}
    \centering
    \includegraphics[width=0.5\linewidth]{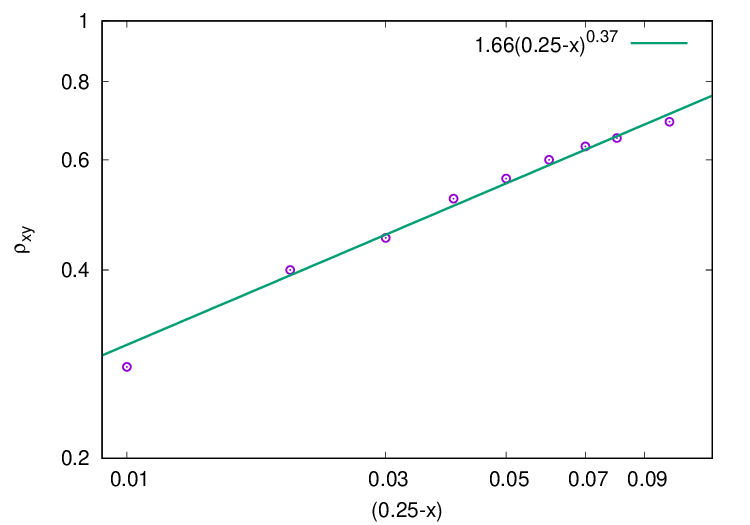}
    \caption{ The variation of correlation coefficients $\rho_{xy}$ is studied as a function of $p_c-p$ for $p < p_c$. The corresponding critical exponent is determined to be $0.37\pm0.02$. }
    \end{figure}
This new  critical exponent thus characterises the shape change of bivariate distribution. 

\begin{figure}
\includegraphics[width=9cm]{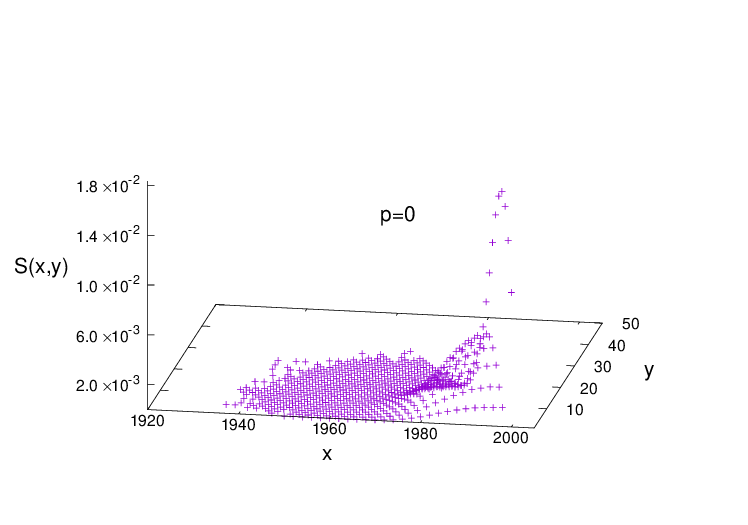}
\caption{The bivariate distribution obtained from simulation data for $p=0$ is shown in the figure for 
positive values of $x$ only.}
\label{bv0}
\end{figure}

As mentioned earlier, for $p=0$, the distributions differ markedly from that for $0< p< p_c$; here  $S(x,y,t)$ cannot be fitted with a continuous bivariate form as it shows several isolated discrete data points deviating from the main branch (see Fig. \ref{bv0}). The detailed results for $p=0$ and $p \neq 0$  are presented separately in the next two subsections.


\subsubsection{Results for $p=0$; $P_{0x}$ and persistence probability}



The marginal probability densities along the X and Y axes are defined according to Eq. \ref{margx} and Eq. \ref{margy} and for $p=0$ are denoted  by $P_{0x}(x,t)$ and $P_{0y}(y,t)$ respectively.
The results are shown in Figures \ref{p0},  \ref{p0y} where it is evident that the distributions are non-Gaussian. 

Further, we have succeeded in getting data collapses for different times (Fig. \ref{p0}b and Fig. \ref{p0y}b).
From Fig. \ref{p0}b, we find that $P_{0x}(x,t)$
has the following form
\begin{equation}
P_{0x}(x,t) \propto\frac{f_0(z_0)}{(\log t)^{0.4}}e^{-\alpha_0 z_0^2},
\label{mgx0}
\end{equation}
where $z_0=\frac{x-\lambda_0 t}{(\log t)^{0.4}}$  and\\
\begin{equation}
 f_0(z_0)\propto 1+b_1\erf(b_2(z_0-b_3)).
 \label{mgcx0}
\end{equation}
On the other hand 
as shown in Fig. \ref{p0y}a, $P_{0y}(y,t)$ can be fitted to the form
\begin{equation}
P_{0y}(y,t)=\frac{g_0(v_0)}{t^{0.23}}e^{-\beta_0 v_0^2},
\label{mgy0}
\end{equation}
where $v_0=\frac{y}-\omega_0 t^{0.97}{t^{0.23}}$ and 
\begin{equation}
    g_0(v_o)\propto [1+k_1\erf(k_2(v_0-k_3))].
    \label{mgcy0}
\end{equation}
$b_i $ and $k_i$ occurring in the scaling functions in equations \ref{mgcx0} and \ref{mgcy0} are constants. As it is apparent, both the forms deviate from a Gaussian form to a large extent. 

\begin{figure}
\includegraphics[width=8cm]{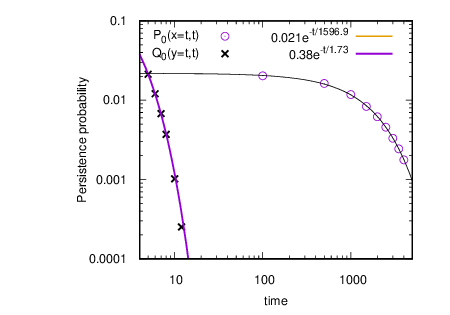}
\caption{For $p=0$, the endpoints of the marginal distributions along the positive X and Y-axes, i.e., $P_{x0}(x=t,t)$ and $P_{y0}(y=t,t)$, which represent the persistence probabilities of $+1$ and $0$ opinions, respectively, are observed to decrease exponentially over time.}
\label{pers0}
\end{figure}

According to the definition of the walking scheme, $P_{0x}(x=t,t) ~~ (=P_{0x}(x=-t,t))$ is the persistence probability for opinion $+1$ ($-1$). Similarly, the persistence probability of 
zero opinion is given by  $P_{0y}(y=t,t)$. However, the scaling forms found above 
are not valid for extreme values of $x,y$ and hence one can obtain the behavior of the persistence probability by directly studying the behavior of  $P_{0x}(x=t,t)$ and  $P_{0y}(y=t,t)$.
In Fig. \ref{pers0} we have shown that  $P_{0x}(x=t,t)$ and $P_{0y}(y=t,t)$ both decay exponentially over time. However, the decay for $P_{0y}$ is much faster which shows that the
zero opinions survive for a much shorter period. This can be understood  as the zero opinion can 
remain zero only when interaction with another agent with opinion zero  takes place. In fact, the results show that the timescales associated with the decay of the persistence probabilities of $\pm 1$ and zero opinions differ by several orders of magnitude.



\subsubsection{Results for $p\neq0$}

\begin{figure}[ht]
    \centering
    \begin{minipage}{0.45\textwidth}
        \vspace{-1.8cm}
        \includegraphics[width=1.35\linewidth]{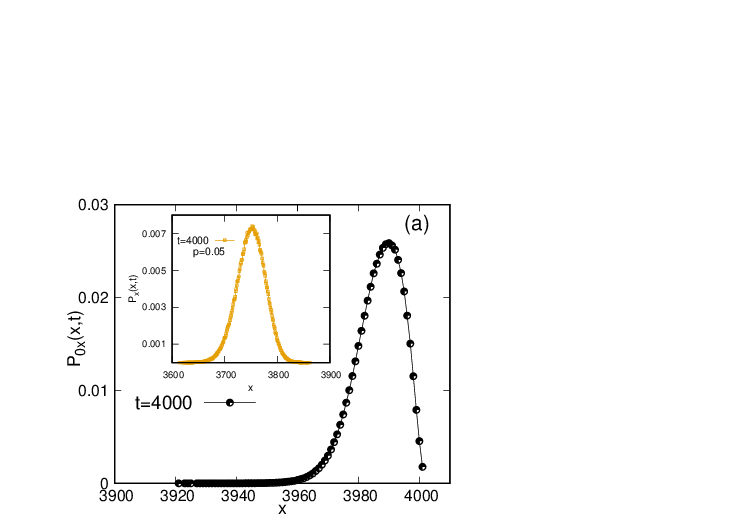}
    \end{minipage}
    \begin{minipage}{0.32\textwidth}
        \vspace{-0.2cm}
        \hspace{-1.65cm}
        \includegraphics[width=1.25\linewidth]{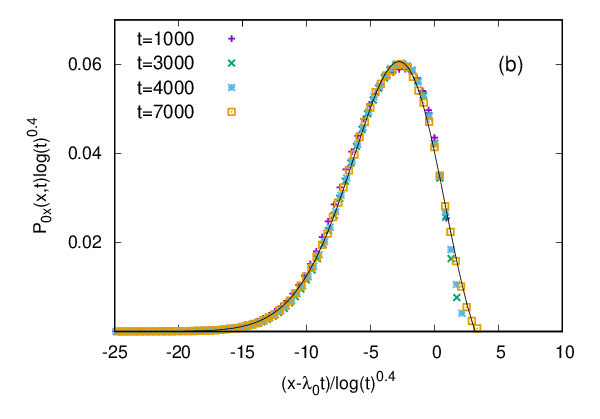}
    \end{minipage}
    \caption{ (a) $P_{0x}(x,t)$ of the walk along the positive X axis for $p=0$ at $t=4000$. A comparison with $p=0.05$ shown in the inset clearly indicates a change in the behaviour of the marginal distribution along the X-axis. A comparison is shown with the data for $p=0.05$ in the inset. (b) Data collapse of $P_{0x}(x,t)(\log t)^{0.4} $ for  different times at $p=0$ and $x>0$ plotted against the scaling variable 
$z_0=\frac{x-\lambda_0t}{(\log t)^{0.4}}$. Where $\lambda_0 \sim 0.999$.} 
 
    \label{p0}
\end{figure}

\begin{figure}[ht]
    \centering
    \begin{minipage}{0.45\textwidth}
        \vspace{-1.0cm}
        \includegraphics[width=1.35\linewidth]{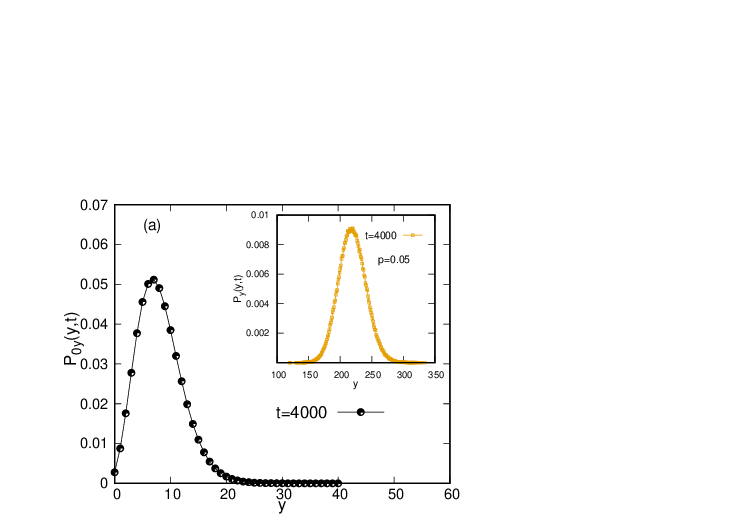}
    \end{minipage}
    \begin{minipage}{0.4\textwidth}
        \vspace{-1.4cm}
        \hspace{-0.8cm}
        \includegraphics[width=1.4\linewidth]{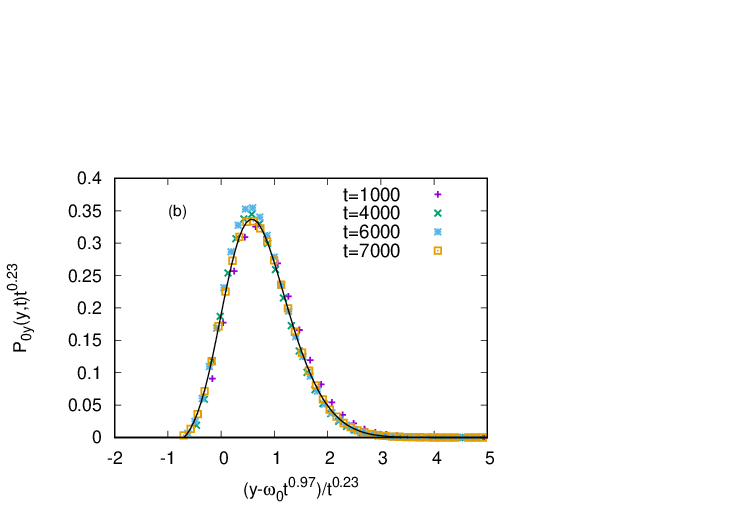}
    \end{minipage}
    \caption{ (a) $P_{0y}(y,t)$ of the walk along the Y axis for $t=4000$. A comparison with $p=0.05$ shown in the inset clearly indicates a change in the behaviour of the marginal distribution along the Y-axis. 
    (b) Data collapse of $P_{0y}(y,t)t^{0.23}$ for different times at $p=0$ plotted against the scaling variable $v_0=\frac{x-\omega_0t^{0.97}}{t^{0.23}}$. Where $\omega_0 \sim 0.001$ . The inset shows the data collapse in the log scale.}
 
    \label{p0y}
\end{figure}

 We have shown in the insets of Figures \ref{p0}a and \ref{p0y}a the data for a small value $p=0.05$, for which the distributions resemble a (biased) Gaussian to a large extent. So, 
 even a small positive value of $p$ can change the nature of the distribution and in fact, we find that for $0<p<p_c$, the form of the distribution is completely different compared to that at $p=0$. 
 
 $P_x(x,t)$ is the marginal probability distribution along the X axis and also identical to the probability distribution of the 1D virtual walk studied earlier \cite{surajit1}.
 Since the $p=0$ results show a clear deviation from a biased Gaussian nature, we explore the possibility of such deviations occurring for non-zero values of $p$, although the figures indicate  that the deviation, if any, for $p \neq 0$ is much less compared to that for $p=0$.  
 In the present work therefore, we have studied the distributions using  a larger number of  configurations  (typically 200000) compared to that in  \cite{surajit1}.

\begin{figure}
\includegraphics[width=0.49\textwidth]{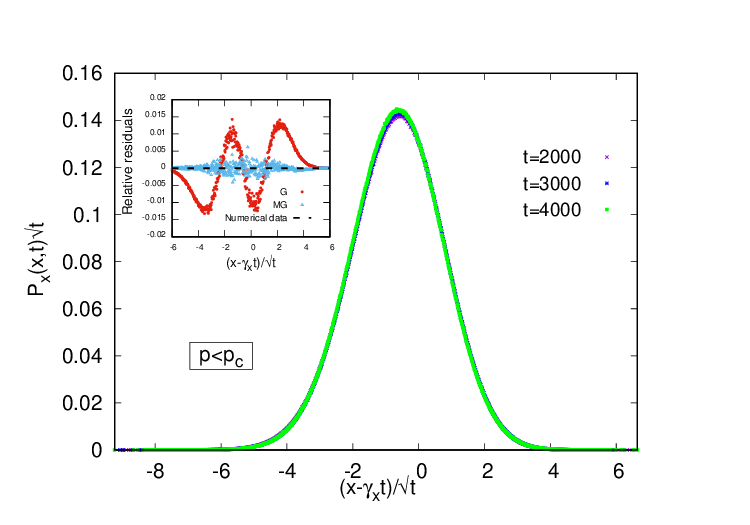}
\caption{Data collapse of $P_x(x,t)\sqrt{t}$ for $p=0.19$ and $x>0$ plotted against the scaling variable 
$z_x=\frac{x-\gamma_x t}{\sqrt{t}}$ where $\gamma_x \sim 0.605$. The inset shows relative residuals for Gaussian (G) fit and a MG (Eq.\ref{mgx}) fit. Relative residuals of the fittings are defined by $\frac{P_x(x,t)\sqrt{t}-(fitted \,\, function)}{P_{x\mid max}(x,t)}$ where $P_{x\mid max}(x,t)$ denotes the maximum value of $P_{x}(x,t)$. The numerical data are shown as a long-dashed black curve and the red and blue dots refer to the G and MG fit respectively.  Non-negligible residuals for G confirm a MG form of $P_x(x,t)$ which leads to a MG marginal distribution along the X-axis.
}
\label{dc}
\end{figure}

 Indeed, we find that using a {\it modified} Gaussian (MG) form, a reasonably good collapse can be obtained for $0<p<p_c$ when $P_x(x,t)\sqrt{t}$ is plotted against $z_x=\frac{(x-\gamma_x t)}{\sqrt{t}}$ shown in Fig. \ref{dc}. 


\begin{figure}
\includegraphics[width=0.49\textwidth]{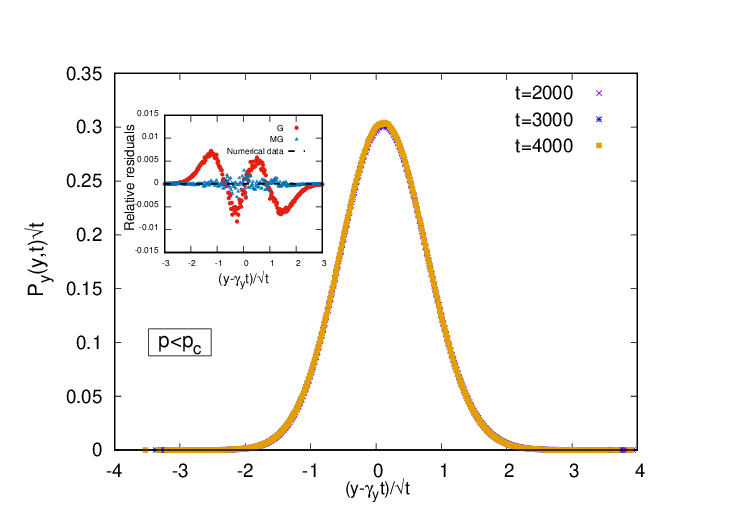}
\caption{Data collapse of $P_y(y,t)\sqrt{t}$ for $p=0.19$  plotted against the scaling variable $z_y=\frac{y-\gamma_y t}{\sqrt{t}}$ is shown , where $\gamma_y \sim 0.235$.  The inset shows relative residuals for Gaussian (G) fit and a MG fit according to the form given in the text. Relative residuals of the fittings are defined by $\frac{P_y(y,t)-(fitted \,\, function)}{P_{y\mid max}(y,t)}$, where $P_{y\mid max}(y,t)$ denotes the maximum value of $P_{y}(y,t)$. The numerical data are shown as a long-dashed black curve and the red and blue dots refer to the G and MG fit respectively.  Non-negligible residuals for G confirm a MG form of $P_y(y,t)$ which leads to a MG marginal distribution along the Y-axis.}
\label{dcy}
\end{figure}

\begin{figure}
\includegraphics[width=0.49\textwidth]{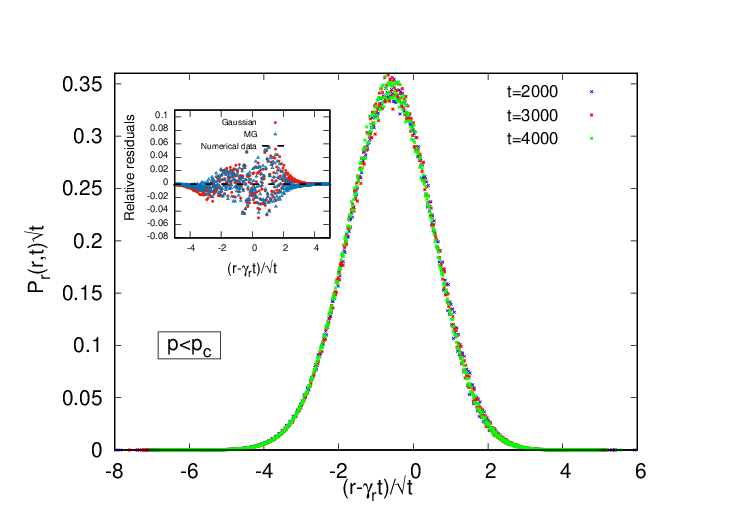}
\caption{Data collapse of $P_r(r,t)\sqrt{t}$ plotted against the scaling variable $z_r=\frac{r-\gamma_rt}{\sqrt{t}}$ is shown for $p=0.19$, where $\gamma_r \sim 0.65$. A Gaussian fit (black line) has been shown which deviates from numerical data. The inset shows relative residuals for Gaussian fit and a MG fit according to the form given in the text. Relative residuals of the fittings are defined by $\frac{P_r(r,t)-(fitted \,\, function)}{P_{r\mid max}(r,t)}$, where $P_{r\mid max}(r,t)$ denotes the maximum value of $P_{r}(r,t)$. }
\label{dcr}
\end{figure}

 The MG form can be written as follows:
\begin{equation}
P_x(x,t)=\frac{f(z_x)}{\sqrt{t}} e^{-\alpha z_x^2}.
\label{mgx}
\end{equation}
Here, both
 $\gamma_x$,  occurring in $z_x$, and $\alpha$ are functions of $p$ and
 \begin{equation}
 f(z_x)\propto 1+w_1 \erf{(w_2(z_x-w_3))},
 \label{fzeq}
 \end{equation}
 which has the same form as  $f_0(z_0)$ given in Eq. \ref{mgcx0} for $p=0$.
 So, the walk deviates from the biased random walk even for nonzero values of $p$ with  $p<p_c$.
Similar types of data collapse can be achieved for  $P_y(y,t)$ (Fig. \ref{dcy}) and $P_r(r,t)$ (Fig. \ref{dcr}). 
Precisely
$P_y(y,t)=\frac{q(z_y)}{\sqrt{t}} e^{-\beta z_y^2}$ 
and $P_r(r,t)=\frac{d(z_r)}{\sqrt{t}} e^{-\beta z_r^2}$ 
where $z_y=\frac{(y-\gamma_y t)}{\sqrt{t}}$ and 
 $z_r=\frac{(r-\gamma_r t)}{\sqrt{t}}$.
 The functions ${q(z_y)}$ and $d(z_r)$ have the same form as ${f(z_x)}$ as given in 
 Eq. \ref{fzeq}.  $\gamma_y$ and $\gamma_z$ are also dependent on $p$.

To establish more strongly that the distributions, $P_x(x,t)$ and $P_y(y,t)$ are not purely Gaussian, we estimated the relative errors using both the MG form and 
the  biased Gaussian form. The relative errors for the biased Gaussian form are significant and exhibit systematic deviations for both $P_x(x,t)$ and $P_y(y,t)$, rather than appearing random. These results are illustrated in the insets of Fig. \ref{dc} and Fig. \ref{dcy}. 
Hence we claim that fitting with a MG improves the precision for these two marginal distributions. 
For $P_r(r,t)$, however, the errors are comparable (Fig. \ref{dcr}).
But it has to be noted here that since $r$ can be non-integer in general,  there is an additional  coarse graining involved for $P_r(r,t)$ when computed numerically. Hence the results are not as precise as that of the other two distributions. 



\begin{figure}
\includegraphics[width=0.49\textwidth]{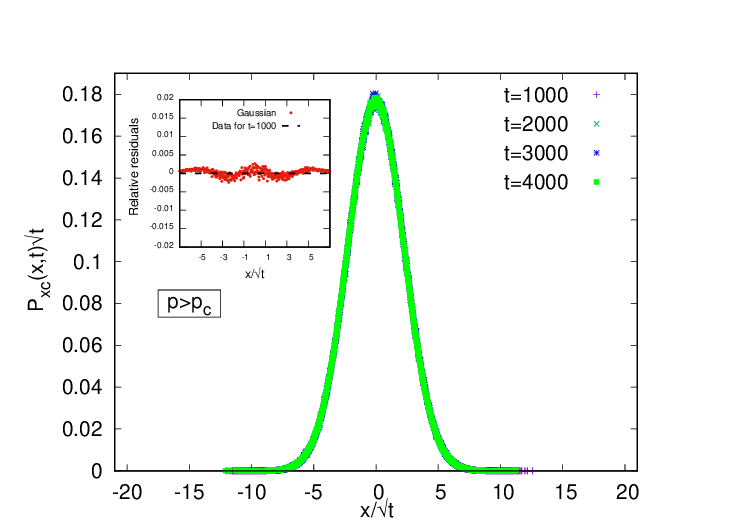}
\caption{Data collapse of $P_{xc}(x,t)\sqrt{t}$ for $p=0.26> p_c$ plotted against 
$\frac{x}{\sqrt{t}}$. A Gaussian fit (black line)
 for $t=1000$ has been shown. The inset
shows the relative residuals of the fittings. The numerical data are shown as a
long-dashed black curve and the red dots refer to the Gaussian fit. residuals are negligible which confirms a Gaussian form of $P_c(x,t)$.}
\label{dcc}
\end{figure}

 Above criticality ($p>p_c$), the marginal distribution along X-axis, written as  $P_{xc}(x,t)$,  behaves like an unbiased random walk.  The relative residuals of the Gaussian fitting are negligible compared to that for below $p_c$. Data collapse of  $P_{xc}(x,t)$ and the
 comparison with the Gaussian fitting have been shown in Fig. \ref{dcc}.   
 One can write  
\begin{equation}
P_{xc}(x,t) \propto  \frac{1}{\sqrt{t}}e^{-\frac{\alpha_c x^2}{t}},
\label{pxx}
\end{equation}
but along the Y-axis the distribution still deviates from Gaussian. 
This is also evident from the deviations of the bivariate distribution for $p>p_c$ from a bivariate normal distribution.





\begin{figure}
\includegraphics[width=0.49\textwidth]{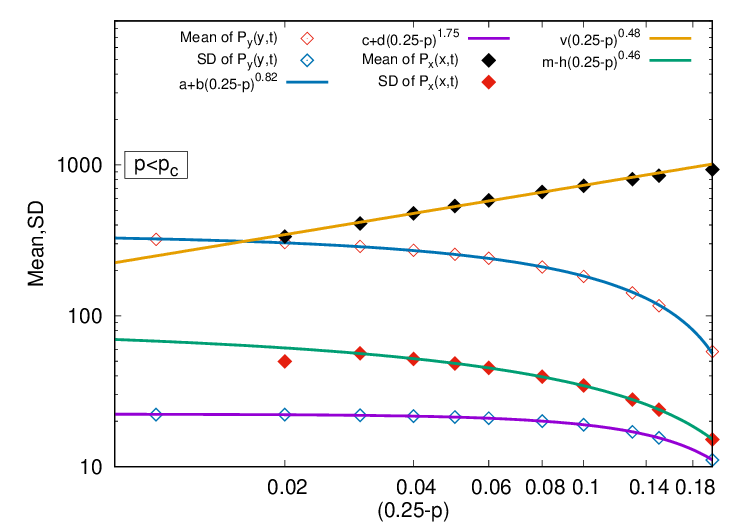}
\caption{Variation of mean and SD of $P_x(x,t)$ and $P_y(x,t)$ against $p_c - p$ for $p<p_c$  and $t=1000$. The values of the critical exponents of Mean and SD for $P_x(x,t)$ are $0.48\pm 0.01$ and $ 0.46\pm 0.03$ respectively.The values of the critical exponents of Mean and SD for $P_y(y,t)$ are $0.82\pm 0.01$ and $1.75\pm 0.01$ respectively. 
 }
\label{bcx}
\end{figure}


\subsection{Moments of the distributions}

Next, we examined the variations of the mean and standard deviations of the marginal distributions for $0<p<p_c$ and $p>p_c$ as a function of $(p_c-p)$. As we approach the critical point ($p_c=0.25$), it is anticipated that the bimodal distribution will approach a unimodal form such that the mean value of $P_x(x,t)$  (Eq. \ref{mgx}) should approach zero, i.e., 
the bias will become zero as given by Eq. \ref{pxx}.
 We note by analyzing the data that  the mean of $P_x(x,t)$ 
 
\begin{equation} 
M_x  \propto(p_c-p)^{\delta_1}. 
\label{ws1}
\end{equation} 
where $\delta_1 \sim 0.48$.
The  standard deviation (SD) shows the behaviour
\begin{equation} 
D_x =m-h(p_c-p)^{\sigma_1}; 
\label{ww1}
\end{equation} 
where $\sigma_1\sim 0.46$ and $m,h$ are constants independent of $p$.
 
 The behavior of the mean and SD of $P_y(x,t)$  have also been obtained. The mean of $P_y(x,t)$  behaves as follows
\begin{equation}
 M_y =a+b(p_c-p)^{\delta_2}; 
\label{wm1}
\end{equation} 
where $\delta_2 \sim 0.82$ and the SD of $P_y(y,t)$
\begin{equation} 
D_y =c+d(p_c-p)^{\sigma_2}; 
\label{w11}
\end{equation} 
where $\sigma_2\sim 1.75$ and $a,b,c,d$ are constants independent of $p$.
 For $p>p_c$, SD and mean become constant as $p$ increases.
 
  All the above data with the best fit curves with  the associated 
 errors are shown in Fig. \ref{bcx}.






\subsection{Diffusion property of the walk}
At $p=0$, the walk is nearly ballistic along X axis as noted in the behavior of the distribution (see Fig. \ref{p0}a): it has a prominent peak very close to $x=t$ and a relatively small width. Indeed,  $\lambda_0$ is very close to one in consistency with this observation. However, the presence of a nonzero width of the distribution for $x>0$ (and $x<0$) clearly signifies a deviation from purely ballistic motion.  We find that for $x>0$,
\begin{equation}
     \sqrt{\langle x^2 \rangle-\langle x \rangle^2}\propto (\log t)^{0.4}.
\end{equation}

\begin{figure}
\includegraphics[width=0.49\textwidth]{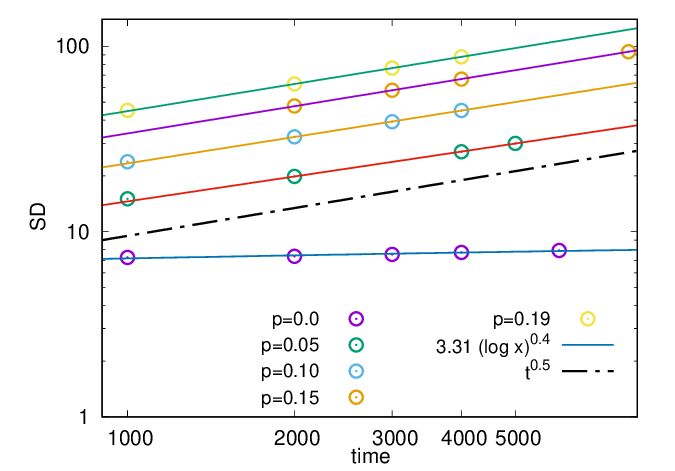}
\caption{ Variation of the SD of $P_x(x,t)$ ($x>0$) over time for different values of $p$ with $p<p_c$.}
\label{dif}
\end{figure}

The origin of such deviations lies in the existence of a small but nonzero probability of taking  steps along the Y axis. For a purely ballistic walk, one expects
 $\langle x^2 \rangle=t^2=\langle x \rangle ^2$. However, our findings indicate that
 \begin{equation}
   \langle x^2 \rangle = k t^2,  
 \end{equation}
where $k<1$, further manifesting the deviation from a ballistic trajectory.
For $0<p\leq p_c$, the marginal walk along the X 
resembles a biased random walk. As one can observe, the deviations from the Gaussian profile are small [Fig. \ref{dc}, \ref{dcy}] and $\langle x^2 \rangle\propto t^2$. Also, the measure of the width for $x>0$ or $x<0$ gives the following time dependence:
\begin{equation}
  \sqrt{\langle x^2 \rangle-\langle x \rangle^2}\propto t^d,   
\end{equation}
where $d$ is equal to 0.5 for all practical purposes [shown in Fig. \ref{dif}].

Hence we conclude that the walk is superdiffusive (nearly ballistic) for $p=0$ and even though
for $0 < p < p_c$ the form of the distribution show significant departure from a biased Gaussian, the scaling behavior suggests a (nearly) diffusive walk. For $p\geq p_c$, the walk is diffusive along X as already discussed.

\section{Conclusions}
 Our main purpose of study of the virtual walk is to see whether any nontrivial property of the system can be revealed using the probability density function of the walkers which can provide new physical insights. A new kind of walk in 2D space has been introduced in the present paper, inspired by an opinion dynamics model.  While such virtual walks, corresponding to discrete-time stochastic models, have previously been studied only in one dimension, the present study explores them in two dimensions for the first time, to the best of our knowledge, incorporates anisotropy as well. The extension to two dimensions naturally reveals a more detailed picture of the microscopic features.
Correlations in interacting particle models are usually defined for higher dimensional systems. Here, we have studied the correlations between two  variables representing the 2D virtual walker's $x$ and $y$ coordinates and found that they are negatively correlated.  An associated critical exponent which characterises the shape of the bivariate distribution of the walker is obtained.

The role of the zero opinion was suppressed in the 1D virtual walks but when viewed in 2D it came out of the compressed state and became visible as marginal distribution along the Y axis. Different critical exponents corresponding to the marginal distributions and bivariate distribution of the walk can define universality class for three-state interacting spin systems. These  can be compared in future with other three-state interacting spin systems to see whether they belong to the same universality class or not. Moreover, one can also obtain a  new 1-D distribution,  $P_r(r,t)$, which can be only calculated from the 2-D walks.\\

  The probability densities of the 1D walk for $0<p<p_c$ appeared to be biased Gaussian in the previous study \cite{surajit1}. Observing that for $p=0$ the behavior of the marginal distributions are far from being Gaussian, we have conducted a more precise study which shows that it is indeed a modified biased Gaussian for $p_c>p>0$ also but the modifications are quite small compared to $p=0$.  Our findings reveal that the incorporation of noise significantly reduces deviations from a biased Gaussian distribution. 
  This is perhaps because that in absence of noise at $p=0$, the only source of randomness is the disordered initial configuration. 

  
 Here, we must mention that it was previously thought that for the discrete-time non-markovian random walk with complete memory of its history known as Elephant Random Walk, the exact solution in both normal and anomalous diffusion regimes is given by a simple Gaussian random walk
propagator. But  \cite{erwn} presents numerical evidence that in the superdiffusion regimes, the propagator is, in general, deviates from Gaussian, leading to many new and interesting insights into the walk. We expect that the insight gained from the present investigation about the model's microscopic features can extend the model's applicability in real-world phenomena.  

 In this work, we have obtained a two-dimensional virtual walk corresponding to the BChS opinion dynamics model on a fully connected network. An interesting direction for future research is to proceed in reverse, starting from a higher-dimensional discrete-time random walk model to construct new opinion dynamics models. For example, higher-dimensional random walks with memory \cite{mar, silva}, which exhibit intriguing diffusion regimes, could be utilized to develop novel opinion dynamics models.

 Lastly, we note that the BChS model was found to belong to the Ising universality class from the steady state results for the mean field case as well as the finite dimensional model. However, although the static critical behavior coincided with the Ising criticality, the dynamical behavior, e.g., in two dimensions, are quite different. When the walk picture is considered, it is not straightforward to conceive of a two dimensional walk for the two state Ising model \cite{review1}.  So, as already mentioned, a comparison of the 2D walks for the BChS model and other three state models would be meaningful  and there is no guarantee that they will be similar.

\section{Acknowledgments}
S.S  acknowledge support by the Council of Scientific and Industrial Research, Government of India, through a CSIR NET fellowship [CSIR JRF
Sanction No. 09/028(1134)/2019-EMR-I] and P.S acknowledge support by the Council of Scientific and Industrial Research, Government of India, through project no.  03/1495/23/EMR-II.


\appendix
 
\section{Equivalent urn model }\label{appendix:urn}
\begin{figure}
    \centering
    \includegraphics[width=0.8\linewidth]{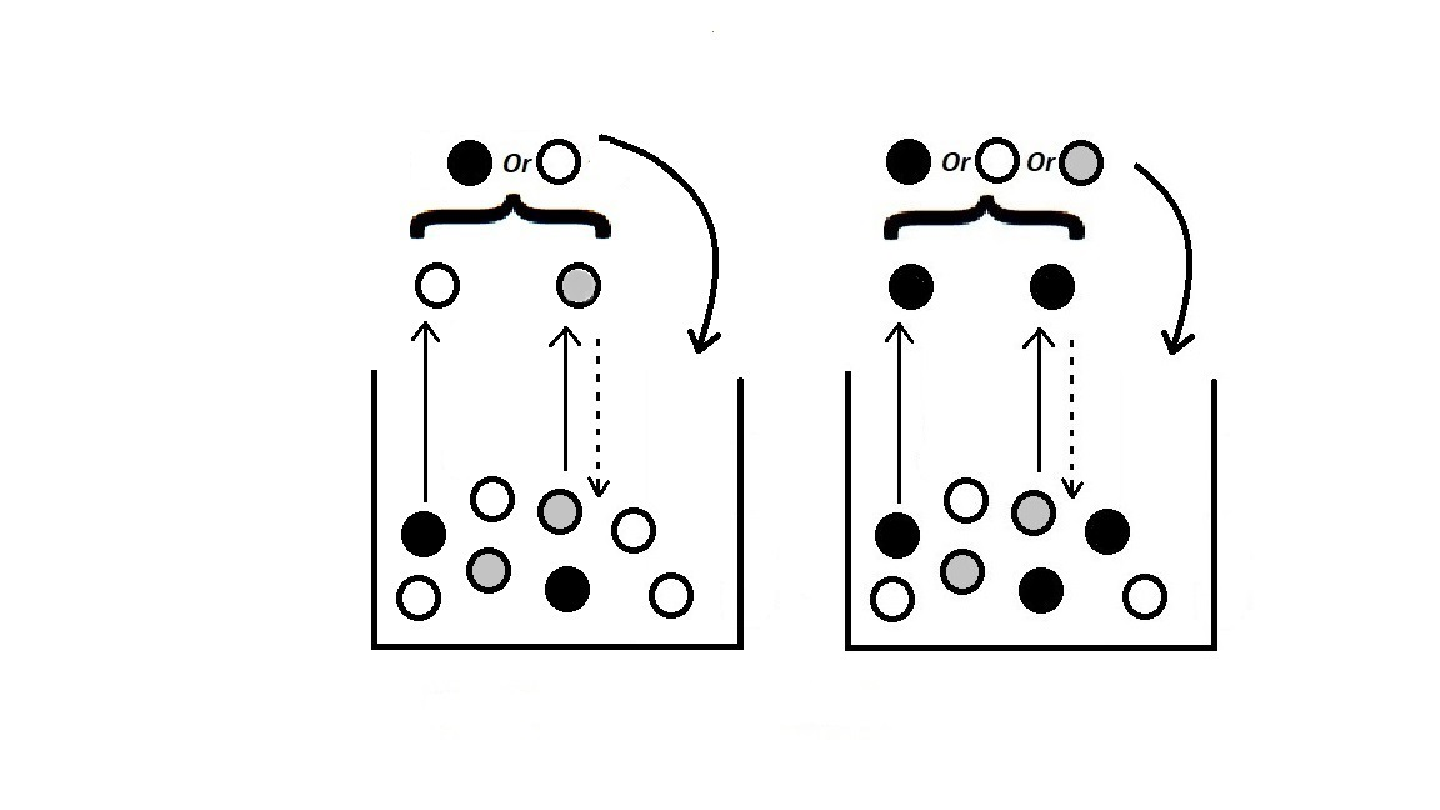}
    \caption{Pólya-type urn containing balls of three different colors
and evolving by drawing two balls at a time according to the mean
replacement matrix (\ref{matrix}). The solid upward straight arrows represent
random drawing of balls in each step, and the dashed downward
arrows represent replacement of the drawn balls. The curved downward arrows denote addition of a new ball according to the mean
replacement matrix (\ref{matrix}.)}
    \label{urn}
\end{figure}

Let us introduce a discrete-time urn model with balls of three colors. Assume the three colors to be black (B), white (W), and gray (G). The composition of the urn at time $t$ is given by a set  $N_t = (N^B_t ,N^W_t,N^G_t)$ where $N^B_t$, $N^W_t$ and $N^G_t$ counts the number of black, white and grey balls respectively. We start at $t=0$ with $N=N^B_0+N^W_0+N^G_0$ balls.
Suppose at each time step two balls were randomly taken out one after the other from the urn (Fig. \ref{urn}). Then the possible outcomes of any drawing can be represented by the following ordered sets: [B, B], [B, W], [W, B], [W, W], [G, W], [W, G], [G, B], [B, G] and [G, G].
After observing the drawn pair, the first ball from it is reinserted into the urn, and a B, W, or G ball is added to it according to rules represented by a mean replacement matrix. So, as time increases, the total number of balls remains constant. The selection of the ball to be added to the urn is determined by the drawn pair, as outlined in the mean replacement matrix below.
 \vspace*{-0.1cm}
\begin{equation}
\begin{blockarray}{cccc}
  &\:\: B & W & G \\
\begin{block}{c(ccc)}
  BB \:\:\,  & a\:& 0\: & b\:  \\
  BW \:\:\,  & b & 0 & a\\
  WB \:\:\,  & 0 & b & a\\
  BG \:\:\,  & 1 & 0 & 0  \\
  GB \:\:\,  & a & b & 0\\
  WW \:\:\,  & 0 & a & b  \\
  WG \:\:\,  & 0 & 1 & 0  \\
  GW \:\:\,  & b & a & 0\\
  GG \:\:\,  & 0 & 0 & 1  \\
 \end{block}
\end{blockarray}
\label{matrix}
\end{equation}
 \vspace*{-0.50cm}
\\
Here $a+b=1$. The elements of the 
matrix represent the conditional probability of the colored ball added in each step. The urn described above belongs to the Pólya type urns evolving by two drawings with randomized replacement rules. If we associate $+1$ to black balls, $-1$ to white balls and $0$ to gray balls then the mean replacement matrix (\ref{matrix}) can be obtained for the choice $a=1-p$ and $b=p$ from Eq.(\ref{op}).\\
 We define $M_t=\frac{\langle N^B_t - N^W_t \rangle}{N}$ as the order parameter of the system.
 One can show that for a particular $p$, 
\begin{eqnarray}
 Number \: of \: +1 \: opinions \:at\: time \: t =_d N^B_t\\ 
 Number \: of \: -1 \: opinions \:at\: time \: t =_d N^W_t\\
 Number \: of \: 0 \: opinions \:at\: time \: t =_d N^G_t
\end{eqnarray}
and,
\begin{equation}
O(t)=_{d} M_t,
\end{equation}
where $=_{d}$ implies equality in distribution. That is to say, the difference between the number of black and white balls in the urn at time $t$ follows the same distribution as the order parameter at time t of the  kinetic exchange model of opinion dynamics defined on a fully connected
graph by the Eq. \ref{op}  with $N$ individuals and  $N^B_0$, $N^W_0$, $N^G_0$ number of $+1$, $-1$ and $0$ individuals respectively at time $t=0$.  
 \\

\begin{figure}
\includegraphics[width=0.4\textwidth]{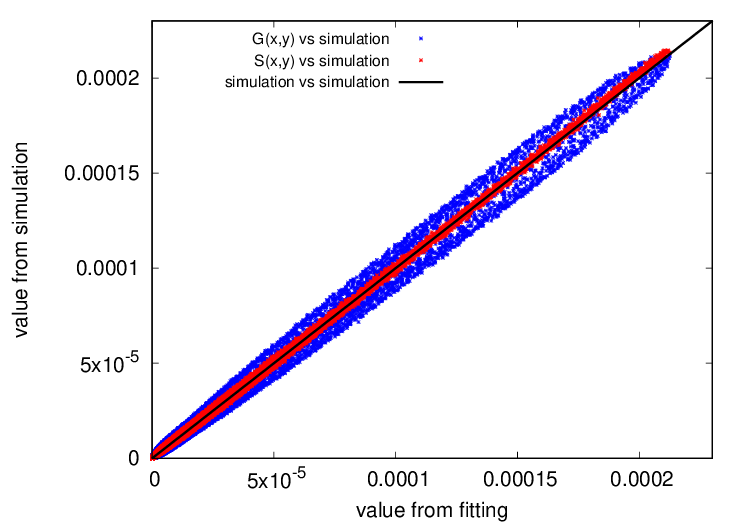}
\caption{q-q plot for the fitting of the bivariate walk distribution at $p=0.19$ with bivariate normal distribution $G(x,y)$ (Eq.\ref{g}) and $S(x,y)$ (Eq.\ref{mg11}). The q-q plot clearly shows that the data for the bivariate distribution $S(x,y)$ (denoted by red dot) is matched to the reference line much better than the bivariate normal distribution $G(x,y)$ (denoted by blue dot).}
\label{qq}
\end{figure}

\begin{figure}
\includegraphics[width=0.4\textwidth]{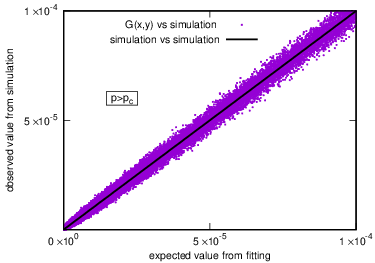}
\caption{q-q plot for the fitting of the bivariate walk distribution at $p=0.26$ with bivariate normal distribution $G(x,y)$ (Eq.\ref{g}). The q-q plot shows that the points for the bivariate normal distribution $G(x,y)$ vs simulation (denoted by blue dot) deviates from the reference line. }
\label{qqc}
\end{figure}

\begin{figure}
\includegraphics[width=8cm]{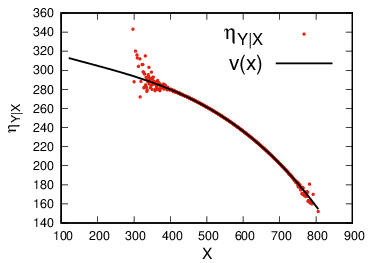}
\caption{ The conditional mean along the $Y$ axis, $\eta_{Y\vert X}$ for $p=0.19$, $N=1000$ and $t=1000$ is fitted with the functional form $v(x)=k_y+l_1(x-k_x)+l_2(x-k_x)^2+l_3(x-k_x)^3$, where $k_y \sim 234.65$ which is very close to $\gamma_y t \sim 235$ (Fig. \ref{dcy}), $k_x \sim \gamma_x t \sim 605$ (Fig. \ref{dc} )and $l_1 \sim -0.296$, $l_2 \sim -4\times10^{-4}$, $l_3 \sim 3.8\times10^{-7}$. }
\label{mx}
\end{figure}
\section{Bivariate distributions and q-q plot}\label{appendix:qq}
For $0<p<p_c$, following modified bivariate normal form of $S(x,y)$ has been confirmed by a quantile-quantile (q-q) plot (Fig. \ref{qq}).
\begin{equation}
S(x,y)=H(x,y)e^{-(m_1(x-\mu_1)^2+m_2(y-\mu_2)^2+m_3(x-\mu_1)(y-\mu_2))}.
\label{mg11}
\end{equation}
Here,\\
$H(x,y)\propto\left\lbrace 1+h_1\erf(h_{2}(x-\mu^{\prime}_1)+h_{3}(y-\mu^{\prime}_{2}))\right\rbrace.$\\
 The q-q plot provides a visual method to assess whether two data sets are derived from populations that share a common distribution. If the two sets come from a population with the same distribution, the points should fall approximately along a 45-degree straight line. The more the deviation from this line, the stronger the indication that the two data sets originate from populations with distinct distributions.
 For example, if we try to fit the observed bivariate distribution from simulation to a bivariate normal distribution with nonzero correlation, i.e.,
 \begin{equation}
G(x,y)=Ae^{-a(x-\nu_x)^2-b(y-\nu_y)^2-c(x-\nu_x)(y-\nu_y)}
 \label{g} 
 \end{equation}
  then the corresponding q-q plot deviates from the 45-degree reference line (Fig. \ref{qq}) but the deviations of the bivariate MG form of Eq. (\ref{mg11}) are much smaller.\\
  The conditional mean along the $Y$ axis, for a given position along the $X$ axis, is defined as follows.
 \begin{equation}
\eta_{Y\vert X}(x,t)=\sum_{y=0}^{\infty}yS(x,y,t)
 \end{equation}
For  $p<p_c$, which can be fitted to the following form for particular $p$ and $t$ (Fig.\ref{mx}).
\begin{equation}
v(x)=k_y+l_1(x-k_x)+l_2(x-k_x)^2+l_3(x-k_x)^3
\label{fx}
\end{equation}
Where $k_x$ and $k_y$ are very close to $\gamma_x t$ and $\gamma_y t$ respectively. In Fig. \ref{mx} we have shown the conditional mean for $p=0.19$. One can obtain a similar kind of behaviour for conditional mean along the X axis also.   It should be noticed that for bivariate normal distribution (Eq.\ref{g}), the conditional mean is a linear function of x 
So, a departure of $\eta_{Y\vert X}$ from the linear dependence on x is also a piece of evidence for the modified normal distributions of $S(x,y,t)$ for $p<p_c$. For $p>p_c$ we have also obtained nonlinear behavior of the conditional mean.   


\end{document}